\PassOptionsToPackage{final}{hyperref}
\PassOptionsToPackage{final}{microtype}
\PassOptionsToPackage{hyphens}{url}

\RequirePackage{fancyvrb}
\RequirePackage[abort, l2tabu, orthodox]{nag}

\documentclass[sigconf,screen]{acmart}

\usepackage{bookmark}
\usepackage{booktabs}
\usepackage{flushend}
\usepackage[final]{listings}
\usepackage{newverbs}
\usepackage{pgfplots}
\usepackage{pgfplotstable}
\usepackage{soul}
\usepackage{subcaption}
\usepackage{tikz}

\usepackage{changes}
\definechangesauthor[name={Jordan}, color={blue}]{j}
\definechangesauthor[name={Ben}, color={red}]{b}

\newcommand{\subsubsubsection}[1]{\smallskip\noindent\textbf{#1.}}
\newcommand{\rqsection}[3]{\section{RQ\ref{rq:#2}: #3}\label{Se:RQ#1}}

\VerbatimFootnotes{}

\usetikzlibrary{arrows}
\usetikzlibrary{arrows.meta}
\usetikzlibrary{positioning}

\usepackage{./sty/trfrac}
\usepackage{./sty/refs}
\usepackage{./sty/rqbox}

\lstdefinelanguage[word2vec]{C}{language=C, morekeywords={assume,call,Called,RetEq,RetNeq,RetLessThan,ParamTo,ParamShare,AccessPathStore,Sensitive,RetConst,PropRet,RetError,Error,FunctionStart,FunctionEnd,AccessPathSensitive}}

\DeclareMathOperator*{\argmax}{arg\,max}

\DeclareMathOperator{\bV}{\mathbb{V}}

\DeclareMathOperator{\cA}{\mathcal{A}}
\DeclareMathOperator{\cB}{\mathcal{B}}
\DeclareMathOperator{\cC}{\mathcal{C}}
\DeclareMathOperator{\cD}{\mathcal{D}}

\DeclareMathOperator{\vERR}{\overrightarrow{\mbox{\$ERR}}}
\DeclareMathOperator{\vEND}{\overrightarrow{\mbox{\$END}}}

\crefname{enumi}{configuration}{configurations}
\creflabelformat{enumi}{#2(#1)#3}

\definecolor{cverbbg}{gray}{0.92}
\sethlcolor{cverbbg}
\newverbcommand{\cverb}
  {\setbox\verbbox\hbox\bgroup}
  {\egroup\colorbox{cverbbg}{\box\verbbox}}

\pgfplotstableset{
  col sep=comma,
}

\newcommand{\addsumcolumn}[1]{%
  \pgfplotstablecreatecol[
  create col/expr={\thisrow{Passed} + \thisrow{Failed} + \thisrow{OOV}}
  ]{sum}{#1}
}

\pgfplotsset{
  compat=1.9,
  percentage nodes near coords/.style={
    nodes near coords=\pgfmathprintnumber{\pgfplotspointmeta}\%,
    nodes near coords bar offset=1,
    every node near coord/.style={
      font=\small,
      /pgf/number format/fixed,
      /pgf/number format/fixed zerofill,
      /pgf/number format/precision=1,
      #1,
    }
  },
  percentage series/.style 2 args={
    table/#1 expr=(\thisrow{#2} / \thisrow{sum} * 100), 
  },
  trinary plot/.style 2 args={
    #2bar stacked,
    #2min=0,
    #2max=100,
    #2majorgrids,
    #2minorgrids,
    #2ticklabel={\pgfmathprintnumber{\tick}\%},
    enlarge #2 limits={
      upper,
      abs value=.2,
    },
    minor #2 tick num=1,
    #1tick=data,
    axis lines*=left,
    typeset ticklabels with strut,
  },
}



\makeatletter
\renewcommand*{\NAT@spacechar}{~}
\makeatother

\title[Code Vectors: Understanding Programs Through EASTs]{Code Vectors: Understanding Programs Through\texorpdfstring{\\}{ }Embedded Abstracted Symbolic Traces}

\begin{document}

\acmBadgeL[https://zenodo.org/badge/latestdoi/122389109]{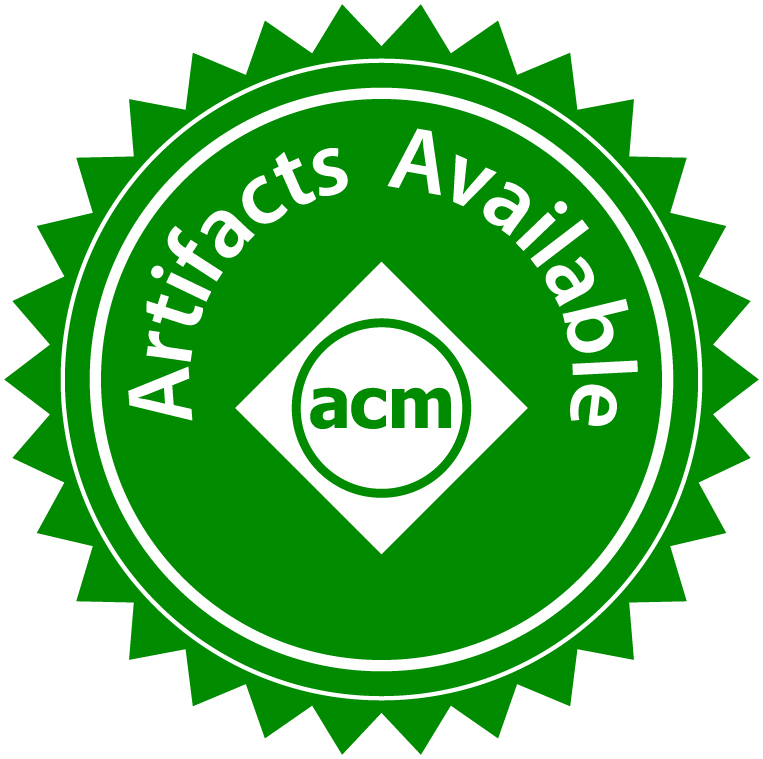}
\acmBadgeR{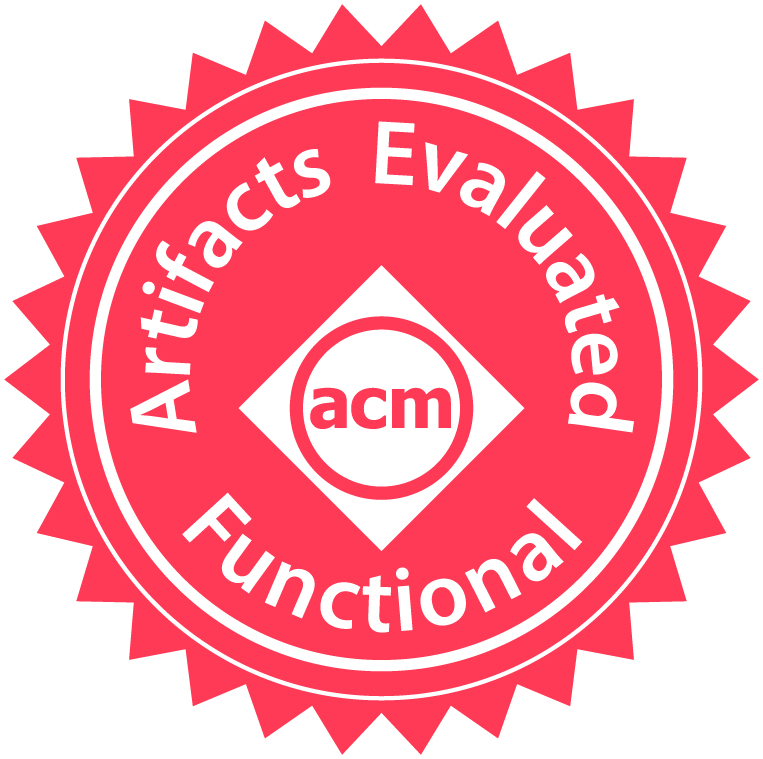}
\settopmatter{printfolios=true}

\copyrightyear{2018} 
\acmYear{2018} 
\setcopyright{acmcopyright}
\acmConference[ESEC/FSE '18]{Proceedings of the 26th ACM Joint European Software Engineering Conference and Symposium on the Foundations of Software Engineering}{November 4--9, 2018}{Lake Buena Vista, FL, USA}
\acmBooktitle{Proceedings of the 26th ACM Joint European Software Engineering Conference and Symposium on the Foundations of Software Engineering (ESEC/FSE '18), November 4--9, 2018, Lake Buena Vista, FL, USA}
\acmPrice{15.00}
\acmDOI{10.1145/3236024.3236085}
\acmISBN{978-1-4503-5573-5/18/11} 

\newcommand{\UWMad}[1][ersity]{Univ#1 of Wisconsin--Madison} 

\author{Jordan Henkel}
\orcid{0000-0003-3862-249X}
\affiliation[obeypunctuation=true]{%
  \institution{\UWMad},
  \country{USA}
}
\email{jjhenkel@cs.wisc.edu}

\author{Shuvendu K. Lahiri}
\affiliation[obeypunctuation=true]{%
  \institution{Microsoft Research},
  \country{USA}
}
\email{Shuvendu.Lahiri@microsoft.com}

\author{Ben Liblit}
\orcid{0000-0002-2245-2839}
\affiliation[obeypunctuation=true]{%
  \institution{\UWMad},
  \country{USA}
}
\email{liblit@cs.wisc.edu}

\author{Thomas Reps}
\orcid{0000-0002-5676-9949}
\affiliation[obeypunctuation=true]{%
  \institution{\UWMad[.] and GrammaTech, Inc.},
  \country{USA}
}
\email{reps@cs.wisc.edu}

\begin{abstract}
With the rise of machine learning, there is a great deal of interest in
treating programs as data to be fed to learning algorithms.
However, programs do not start off in a form that is immediately
amenable to most off-the-shelf learning techniques.
Instead, it is necessary to transform the program to a suitable
representation before a learning technique can be applied.

In this paper, we use abstractions of traces obtained from symbolic execution
of a program as a representation for learning word embeddings.
We trained a variety of word embeddings under hundreds of parameterizations,
and evaluated each learned embedding on a suite of different tasks.
In our 
evaluation, we obtain 93\% top-1 accuracy on a benchmark consisting of over 
19,000 API-usage analogies extracted from the Linux kernel. In addition, 
we show that embeddings learned from (mainly) semantic abstractions provide 
nearly triple the accuracy of those learned from (mainly) syntactic abstractions. 


\end{abstract}


\maketitle

\renewcommand{\shortauthors}{J. Henkel, S. Lahiri, B. Liblit, and T. Reps}

\section{Introduction\label{Se:INTRO}}

Computer science has a long history of considering programs as data
objects~\citep{INRIA:DGHKL80,Proc:PADO85}.
With the rise of machine learning, there has been renewed interest in
treating programs as data to be fed to learning algorithms~\citep{DBLP:journals/corr/abs-1709-06182}.
However, programs have special characteristics, including several
layers of structure, such as a program's context-free syntactic
structure, non-context-free name and type constraints, and the
program's semantics.
Consequently, programs do not start off in a form that is immediately
amenable to most off-the-shelf learning techniques.
Instead, it is necessary to transform the program to a suitable
representation before a learning technique can be applied.

This paper contributes to the study of such representations in the context of
\emph{word embeddings}.
Word embeddings are a well-studied method for converting a corpus of natural-language
text to vector representations of words embedded into a low-dimensional space.
These techniques have been applied successfully to programs
before~\citep{Nguyen2017,Pradel2017,Gu:2016:DAL:2950290.2950334},
but different encodings of programs into word sequences are possible, and some encodings may
be more appropriate than others as the input to a word-vector learner.

The high-level goals of our work can be stated as follows:

\begin{focusbox}
  Devise a parametric encoding of programs into word sequences that
  (i) can be tuned to capture different representation choices
  on the spectrum from (mainly) syntactic to (mainly) semantic,
  (ii) is amenable to word-vector-learning techniques, and
  (iii) can be obtained from programs efficiently.
\end{focusbox}

\smallskip
\noindent
We also wish to understand the advantages and disadvantages of our
encoding method.
\sectref{RQ1}--\sectref{RQ4} summarize the experiments that we performed
to provide insight on high-level goal (ii).

We satisfy high-level goals (i) and (iii) by basing the
encoding on a lightweight form of \emph{intraprocedural symbolic execution}.
\begin{itemize}
  \item
    We base our technique on \emph{symbolic execution} due to the gap between syntax
    (e.g., tokens or abstract syntax trees (ASTs)) and the semantics of a procedure
    in a program.
    In particular, token-based techniques impose a heavy burden on the embedding
    learner.
    For instance, it is difficult to encode the difference between
    constructions such as \cverb|a == b| and
    \cverb|!(a != b)| 
    via a
    learned, low-dimensional embedding~\citep{Allamanis2016a}.
  \item
    Our method is \emph{intraprocedural} so that different procedures can be
    processed in parallel.
  \item
    Our method is \emph{parametric} in the sense that we introduce a level of
    mapping from symbolic-execution traces to the word sequences that are input
    to the word-vector learner.
    (We call these \emph{abstraction mappings} or \emph{abstractions},
    although strictly speaking they are not abstractions in the sense
    of abstract interpretation~\citep{POPL:CC77}.)
    Different abstraction mappings can be used to extract different
    word sequences that are in different positions on the spectrum of
    (mainly) syntactic to (mainly) semantic.
\end{itemize}
We have developed a highly parallelizable toolchain that is capable of
producing a parametric encoding of programs to word sequences.  For
instance, we can process 311,670 procedures in the Linux
kernel\footnote{Specifically, we used a prerelease of Linux 4.3
  corresponding to commit
  \href{https://github.com/torvalds/linux/tree/fd7cd061adcf5f7503515ba52b6a724642a839c8}{\texttt{fd7cd061adcf5f7503515ba52b6a724642a839c8}}
  in the GitHub Linux kernel repository.} in 4 hours,\footnote{During
  trace generation, we exclude only \cverb|vhash_update|, from
  \texttt{crypto/vmac.c}, due to its size.} using a 64-core
workstation (4 CPUs each clocked at 2.3 GHz) running CentOS 7.4 with
252 GB of RAM\@.

After we present our infrastructure for generating parametric encodings of
programs as word sequences (\sectref{Overview}), there are a number of natural
research questions that we consider.

First, we explore the utility of embeddings learned from our toolchain:

\begin{rqbox}{useful}
  Are vectors learned from abstracted symbolic 
  traces encoding \textit{useful} information?
\end{rqbox}

Judging utility is a difficult endeavor.
Natural-language embeddings have the
advantage of being compatible with several canonical benchmarks
for word-similarity prediction or analogy solving~\citep{the-microsoft-research-sentence-completion-challenge,
Finkelstein:2001:PSC:371920.372094,Luong2013BetterWR,Szumlanski,
Hill:2015:SES:2893320.2893324,Rubenstein:1965:CCS:365628.365657,
NIPS2013_5021}.
In the domain of program understanding, no such canonical benchmarks exist.
Therefore, we designed a suite of over nineteen thousand code
analogies to aid in the evaluation of our learned vectors.

Next, we examine the impact of different parameterizations of our 
toolchain by performing an ablation study. The purpose of this study is to 
answer the following question:

\begin{rqbox}{best}
  Which abstractions produce the best program
  encodings for word-vector learning?
\end{rqbox}

There are several examples of learning from syntactic artifacts, such as ASTs or
tokens. The success of such techniques raises the question of whether adding a
symbolic-execution engine to the toolchain improves the quality of
our learned representations.

\begin{rqbox}{svs}
  Do abstracted symbolic traces
  at the semantic end of the spectrum provide more utility as
  the input to a word-vector-learning technique compared
  to ones at the syntactic end of the spectrum?
\end{rqbox}

Because our suite of analogies is only a proxy for utility in more complex
downstream tasks that use learned embeddings, we pose one more question:

\begin{rqbox}{downstream}
  Can we use pre-trained word-vector embeddings
  on a downstream task?
\end{rqbox}

The contributions of our work can be summarized as follows: 

\textbf{We created a toolchain} for taking a program or corpus of programs
and producing intraprocedural symbolic traces.
The toolchain is based on Docker containers, is parametric, and
operates in a massively parallel manner.
Our symbolic-execution engine prioritizes the
amount of data generated over the precision of the analysis:
in particular, no feasibility checking is
performed, and no memory model is used during symbolic execution.

\textbf{We generated several datasets} of abstracted symbolic traces
from the Linux kernel.
These datasets feature different parameterizations (abstractions),
and are stored in a format suitable for off-the-shelf word-vector learners.

\textbf{We created a benchmark suite} of over 19,000 API-usage analogies.

\textbf{We report on several experiments} using these datasets:
\begin{itemize}
  \item In \sectref{RQ1}, we achieve 93\% top-1 accuracy on a suite of over 
  19,000 analogies.
  \item In \sectref{RQ2}, we perform an ablation study to assess the effects of 
  different abstractions on the learned vectors.
  \item In \sectref{RQ3}, we demonstrate how vectors learned from 
  (mainly) semantic abstractions can provide nearly triple the accuracy of 
  vectors learned from (mainly) syntactic abstractions. 
  \item In \sectref{RQ4}, we learn a model of a specific program behavior (which error a trace is likely to return), and 
  apply the model in a case study to confirm actual bugs found via traditional static analysis.
\end{itemize}

Our toolchain, pre-trained word embeddings, and code-analogy suite
are all available as part of the artifact accompanying this paper; details are given 
in \sectref{Artifact}.

\paragraph{Organization.}
The remainder of the paper is organized as follows:
\sectref{Overview} provides an overview of our toolchain and applications.
\sectref{ABS} details the parametric aspect of our toolchain
and the abstractions we use throughout the remainder of the paper.
\sectref{WV} briefly describes word-vector learning.
\sectref{RQ1}--\sectref{RQ4} address our four research questions.
\sectref{THREATS} considers threats to the validity
of our approach. \sectref{RW} discusses related work. \sectref{Artifact} describes
supporting materials that are intended to help others build on our
work. \sectref{CONC} concludes.



\section{Overview}%
\label{Se:Overview}

Our toolchain consists of three phases: transformation, abstraction, and learning. 
As input, the toolchain expects a corpus of buildable C projects, a
description of abstractions to use, and a word-vector learner. As output, 
the toolchain produces an embedding of abstract tokens to double-precision vectors with 
a fixed, user-supplied, dimension. 
We illustrate this process as applied to the example in \figref{OverviewProg}.

\begin{figure}
  \centering
  \begin{tabular}{c}
\begin{lstlisting}
int example() {
  buf = alloc(12);
  if (buf != 0) {
    bar(buf);
    free(buf);
    return 0;
  } else {
    return -ENOMEM;
  }
}
\end{lstlisting}
  \end{tabular}
  \caption{An example procedure\label{Fi:OverviewProg}}
\end{figure}

\newcommand{\headingtext}{}
\newcommand{\phasesubsection}[1]{%
  \refstepcounter{subsection}
  \renewcommand{\headingtext}{Phase \Roman{subsection}\texorpdfstring{\@}{}: #1}
  \subsubsubsection{\headingtext}
  \addcontentsline{toc}{subsection}{\headingtext}
}

\phasesubsection{Transformation} The first phase of the toolchain enumerates all paths in each source
procedure.  We begin by unrolling (and truncating) each loop so that its body 
is executed zero or one time, thereby
making each procedure loop-free at the cost of discarding many
feasible traces.  We then apply an intraprocedural symbolic executor
to each procedure.  \figref{OverviewTraces} shows the results of this
process as applied to the example code in \figref{OverviewProg}.

\begin{figure}
  \subcaptionbox{Trace 1}{
\begin{lstlisting}
call alloc(12);
assume alloc(12) != 0;
call bar(alloc(12));
call free(alloc(12));
return 0;
\end{lstlisting}}
  \hfill
  \subcaptionbox{Trace 2\label{Fi:OverviewTrace2}}{
\begin{lstlisting}[showlines]
call alloc(12);
assume alloc(12) == 0;
return -ENOMEM;


\end{lstlisting}}
  \caption{Traces from the symbolic execution of the procedure in \figref{OverviewProg}\label{Fi:OverviewTraces}}
\end{figure}

\phasesubsection{Abstraction} Given a user-defined set of abstractions, the second phase of our
toolchain leverages the information gleaned from symbolic execution to
generate abstracted traces. One key advantage of performing some kind
of abstraction is a drastic reduction in the number of possible tokens
that appear in the traces. Consider the transformed
trace in \figref{OverviewTrace2}.  If we want to understand the
relationship between allocators and certain error codes, then we might
abstract procedure calls as \textit{parameterized tokens} of the form
\lstinline[mathescape]{Called($\mathit{callee}$)}; comparisons of
returned values to constants as parameterized
\lstinline[mathescape]{RetEq($\mathit{callee}$, $\mathit{value}$)}
tokens; and returned error codes as parameterized
\lstinline[mathescape]{RetError($\mathit{code}$)} tokens.
\figref{OverviewAbstractions} shows the result of applying these
abstractions to the traces from \figref{OverviewTraces}.

\begin{figure}
  \hfill
  \subcaptionbox{Abstracted Trace 1}{
\begin{lstlisting}
Called(alloc)
RetNeq(alloc, 0)
Called(bar)
Called(free)
\end{lstlisting}
  }
  \hfill
  \subcaptionbox{Abstracted Trace 2\label{Fi:OverviewAbstractions2}}[\widthof{\textbf{(b) Abstracted Trace 2}}]{%
\begin{lstlisting}[showlines]
Called(alloc)
RetEq(alloc, 0)
RetError(ENOMEM)

\end{lstlisting}
  }
  \hspace*{\fill}
  \caption{Result of abstracting the two traces in \figref{OverviewTrace2}\label{Fi:OverviewAbstractions}}
\end{figure}

\phasesubsection{Learning} Our abstracted representation discards irrelevant details, flattens
control flows into sequential traces, and exposes key properties in
the form of parameterized tokens that capture domain information such
as Linux error codes.  These qualities make abstracted traces suitable
for use with a word-vector learner.  Word-vector learners place words
that appear in similar contexts close together in an embedding space.
When applied to natural language, learned embeddings can answer
questions such as ``king is to queen as man is to what?'' (Answer: woman.)
Our goal is to learn embeddings that can answer questions such as:

\begin{itemize}
\item If a lock acquired by calling \lstinline{spin_lock} is released
  by calling \lstinline{spin_unlock}, then how should I
  release a lock acquired by calling \lstinline{mutex_lock_nested}?
  That is, \lstinline{Called(spin_lock)} is to
  \lstinline{Called(spin_unlock)} as
  \lstinline{Called(mutex_lock_nested)} is to what? (Answer: \lstinline{Called(mutex_unlock)}.)
\item Which error code is most commonly used to report allocation
  failures? That is, which
  \lstinline[mathescape]{RetError($\mathit{code}$)} is most related to
  \lstinline{RetEq(alloc, 0)}? (Answer: \lstinline{RetError(ENOMEM)}.)
\item Which procedures and checks are most related to \lstinline{alloc}? (Answers: \lstinline{Called(free)}, \lstinline{RetNeq(alloc, 0)}, etc.)
\end{itemize}

The remainder of the paper describes a framework of abstractions and a
methodology of learning embeddings that can effectively solve these problems. Along the way, 
we detail the challenges that arise in applying word embeddings to abstract
path-sensitive artifacts.



\section{Abstractions\label{Se:ABS}}

\begin{figure*}
  \setlength{\columnwidth}{\textwidth}
  \begin{gather*}
    \trfrac{
      \text{\lstinline{call foo()}}
    }{
      \text{\lstinline{Called(foo)}}
    }
    \qquad
    \trfrac{
      \begin{trgather}
        \text{\lstinline{call bar(foo())}}
      \end{trgather}  
    }{
      \text{\lstinline{ParamTo(bar, foo)}}
    }
    \qquad
    \trfrac{
      \begin{trgather}
        \text{\lstinline{call foo(obj)}}\\[-4pt]
        \text{\lstinline{call bar(obj)}}
      \end{trgather}  
    }{
      \text{\lstinline{ParamShare(bar, foo)}}
    }
    \qquad
    \trfrac{
      \begin{trgather}
        \text{\lstinline{assume foo() == 0}}
      \end{trgather}
    }{
      \text{\lstinline{RetEq(foo, 0)}}
    }
    \qquad
    \trfrac{
      \begin{trgather}
        \text{\lstinline{obj->foo.bar = baz}}
      \end{trgather}
    }{
      \text{\lstinline{AccessPathStore(->foo.bar)}}
    }\\
    \trfrac{
      \text{\lstinline{return -C}} \;\land\; \text{C} \in \text{\lstinline{ERR_CODES}} 
    }{
      \text{\lstinline{RetError(ERR_CODES[C])}, \lstinline{Error}}
    }
    \qquad
    \trfrac{
      \text{\lstinline{return C}} \;\land\; \text{C} \not\in \text{ERR\_CODES} 
    }{
      \text{\lstinline{RetConst(C)}}
    }
    \qquad
    \trfrac{
      \begin{trgather}
        \text{\lstinline{return foo()}}
      \end{trgather}
    }{
      \text{\lstinline{PropRet(foo)}}
    }
    \qquad
    \trfrac{
      \text{\lstinline{PropRet(PTR_ERR)}}
    }{
      \text{\lstinline{Error}}
    }
  \end{gather*}
  \setlength{\belowcaptionskip}{-10pt}
  \caption{Example derivations for selected abstractions\label{Fi:ABS}}
\end{figure*}


One difference between learning from programs and learning from natural language
is the size of the vocabulary in each domain. In natural language, vocabulary
size is bounded (e.g., by the words in a dictionary, ignoring issues like
misspellings). In programs, the vocabulary is essentially unlimited: due to
identifier names, there are a huge number of distinct words that can occur in a
program. To address the issue of vocabulary size, we perform an abstraction
operation on symbolic traces, so that we work with abstracted symbolic traces
when learning word vectors from programs.

\subsection{Abstracted Symbolic Traces}

\sloppy

We now introduce the set of abstractions that we use to create abstracted
symbolic traces. Selected abstractions appear in the conclusions of the
deduction rules shown in \figref{ABS}. The abstractions fall into a few simple
categories. The \lstinline[mathescape]{Called($\mathit{callee}$)} and
\lstinline[mathescape]{AccessPathStore($\mathit{path}$)} abstractions can be
thought of as ``events'' that occur during a trace. Abstractions like
\lstinline[mathescape]{RetEq($\mathit{callee}$, $\mathit{value}$)} and
\lstinline{Error} serve to encode the ``status'' of the current trace: they
provide contextual information that can modify the meaning of an ``event''
observed later in the trace. Near the end of the trace, the
\lstinline[mathescape]{RetError($\mathit{code}$)},
\lstinline[mathescape]{RetConst($\mathit{value}$)}, and
\lstinline[mathescape]{PropRet($\mathit{callee}$)} abstractions provide
information about the result returned at the end of the trace. Taken together,
these different pieces of information abstract the trace; however, the
abstracted trace is still a relatively rich digest of the trace's behavior.

\fussy

With the abstractions described above, we found that the learned vectors were
sub-optimal. Our investigation revealed that some of the properties we hoped
would be learned required leveraging contextual information that was outside the
``window'' that a word-vector learner was capable of observing. For example, to
understand the relationship between a pair of functions like \lstinline{lock}
and \lstinline{unlock}, a word-vector learner must be able to cope with an
arbitrary number of words occurring between the functions of interest. Such
distances are a problem, because lengthening the history given to a word-vector
learner also increases the computational resources necessary to learn good
vectors. 

Due to the impracticality of increasing the context given to a word-vector
learner, we introduced two additional abstractions: \lstinline{ParamTo} and
\lstinline{ParamShare}.
These abstractions encode the flow of data in the trace to make
relevant contextual information available without the need for
arbitrarily large contexts.
As shown in \sectref{RQ2}, the abstractions that encode semantic information,
such as dataflow facts, end up adding the most utility to our corpus of
abstracted traces.
This observation is in line with the results of
\citet{DBLP:journals/corr/abs-1711-00740}, who found that dataflow
edges positively impact the performance of a learned model on
downstream tasks.


\begin{figure}
\begin{lstlisting}
		  // 1,2 FunctionStart
lock(&obj->lock); // 1,2 Call(lock)
foo = alloc(12);  // 1,2 Call(alloc)
if (foo != 0) {   // 1 RetNeq(alloc, 0)
  obj->baz =      // 1 AccessPathStore(->baz)
    bar(foo);     // 1 ParamTo(bar, alloc)
		  // 1 Call(bar)
} else {          // 2 RetEq(alloc, 0)
  unlock(         // 2 ParamShare(unlock, lock)
    &obj->lock);  // 2 Call(unlock)
  return -ENOMEM; // 2 RetError(ENOMEM)
		  // 2 Error
}                 // 2 FunctionEnd
unlock(           // 1 ParamShare(unlock, lock)
  &obj->lock);    // 1 Call(unlock)
return 0;         // 1 RetConst(0)
		  // 1 FunctionEnd
\end{lstlisting}
  \setlength{\belowcaptionskip}{-15pt}
  \caption{Sample procedure with generated abstractions shown as comments\label{Fi:SPROC}}
\end{figure}


We augment the abstractions shown in \figref{ABS}, with the following
additional abstractions, which are similar to the ones discussed above:
\begin{itemize}
  \item
    \lstinline[mathescape]{RetNeq($\mathit{callee}$, $\mathit{value}$)}, \lstinline[mathescape]{RetLessThan($\mathit{callee}$, $\mathit{value}$)}, and others are variants of the
    \lstinline[mathescape]{RetEq($\mathit{callee}$, $\mathit{value}$)} abstraction shown in \figref{ABS}.
  \item
    \lstinline{FunctionStart} and \lstinline{FunctionEnd} are abstractions introduced at the
     beginning and end of each abstracted trace.
  \item
    \lstinline[mathescape]{AccessPathSensitive($\mathit{path}$)} is similar to \lstinline[mathescape]{AccessPathStore};
    it encodes any complex field and array accesses that occur in \lstinline{assume}
     statements.
\end{itemize}

\subsection{Encoding Abstractions as Words}

We now turn to how the encoding of these abstractions as words and sentences
(to form our trace corpus) can impact the utility of learned vectors.
To aid the reader's understanding, we use a sample procedure and describe
an end-to-end application of our abstractions and encodings.

\figref{SPROC} shows a sample procedure along with its corresponding abstractions.
The number(s) 
before each abstraction signify which of the two paths through
the procedure the abstraction belongs to.
To encode these abstractions as words, we need to make careful choices as to what
pieces of information are worthy of being represented as words,
and how this delineation affects the questions we can answer using the learned vectors. 

For instance, consider the \lstinline{RetNeq(alloc, 0)} abstraction.
There are several simple ways to encode this information as a sequence of words: 

\begin{enumerate}
  \item \lstinline{RetNeq(alloc, 0)} $\Longrightarrow$ \cverb|alloc|, \cverb|$NEQ|,
  \cverb|0|
  \item \lstinline{RetNeq(alloc, 0)} $\Longrightarrow$ \cverb|alloc|, \cverb|$NEQ_0|
  \item \lstinline{RetNeq(alloc, 0)} $\Longrightarrow$ \cverb|alloc_$NEQ|, \cverb|0|
  \item \lstinline{RetNeq(alloc, 0)} $\Longrightarrow$ \cverb|alloc_$NEQ_0|
\end{enumerate}

Each of these four encodings comes with a different trade-off.
The first encoding splits the abstraction into several fine-grained words,
which, in turn, reduces the size of the overall vocabulary.
This approach may benefit the learned vectors because smaller vocabularies
can be easier to work with.
On the other hand, splitting the information encoded in this abstraction into
several words makes some questions more difficult to ask.
For example, it is much easier to ask what is most related to
\lstinline{alloc} being not equal to zero when we have just a single word,
\cverb|alloc_$NEQ_0|, to capture such a scenario. 

In our implementation, we use the fourth option.
It proved difficult to ask interesting questions when the abstractions were broken
down into fine-grained words.
This decision did come with the cost of a larger vocabulary.\footnote{We mitigate
  the increase in vocabulary size from constructions like \cverb|alloc_$NEQ_0|
  by restricting the constants we look for.
  Our final implementation only looks for comparisons to constants in the set
  $\{-2,-1,0,1,2,3,4,8,16,32,64\}$.
}
Encodings for the rest of our abstractions are shown in \figref{ENCODS}.\footnote{Because it is not possible to have
\lstinline{ParamShare(X, Y)} or \lstinline{ParamTo(X, Y)} without a \lstinline{Called(X)} following 
them, the abstractions \lstinline{ParamShare(X, Y)} and \lstinline{ParamTo(X, Y)} are encoded as \cverb|Y|
to avoid duplicating \cverb|X|.}
The sentences generated by applying these encodings to \figref{SPROC}
are shown in \figref{SENTS}.


\begin{figure}[tp]
\begin{lstlisting}[language=Caml]
  match abstraction with 
  | Called (x)       -> x
  | ParamTo (_,x)    -> x
  | ParamShare (_,x) -> x
  | RetEq (x,c)      -> x ^ "_$EQ_" ^ c
  | RetNeq (x,c)     -> x ^ "_$NEQ_" ^ c
  (* ... *)
  | PropRet (x)      -> "$RET_" ^ x
  | RetConst (c)     -> "$RET_" ^ c 
  | RetError (e)     -> "$RET_" ^ ERR_CODES[e]
  | FunctionStart    -> "$START"
  | FunctionEnd      -> "$END"
  | Error            -> "$ERR"
  | AccessPathStore (p)     -> "!" ^ p 
  | AccessPathSensitive (p) -> "?" ^ p
\end{lstlisting}
\caption{Encoding of abstractions \label{Fi:ENCODS}}
\end{figure}
  


\begin{figure}
  \SaveVerb{traceboxA}|$START lock alloc alloc_$NEQ_0 !->baz| 
  \SaveVerb{traceboxB}|alloc bar lock unlock $RET_0 $END|
  \subcaptionbox{Trace 1}{%
    \begin{tabular}{ll}
      \colorbox{cverbbg}{\UseVerb{traceboxA}} \\
      \colorbox{cverbbg}{\UseVerb{traceboxB}}
    \end{tabular}}

  \vspace{\baselineskip}

  \SaveVerb{traceboxA}|$START lock alloc alloc_$EQ_0 lock|
  \SaveVerb{traceboxB}|unlock $ERR $RET_ENOMEM $END|
  \subcaptionbox{Trace 2}{%
    \begin{tabular}{ll}
      \colorbox{cverbbg}{\UseVerb{traceboxA}} \\
      \colorbox{cverbbg}{\UseVerb{traceboxB}}
    \end{tabular}}

    \setlength{\belowcaptionskip}{-10pt}
    \caption{Traces for \figref{SPROC} generated by the encoding from \figref{ENCODS}\label{Fi:SENTS}}
\end{figure}





\section{Word2Vec\label{Se:WV}}

Word2Vec is a popular method for taking words and embedding them into a
low-dimensional vector space~\citep{NIPS2013_5021}.
Instead of using a one-hot encoding---where each element of a vector is
associated with exactly one word---word2vec learns a denser representation
that captures meaningful syntactic and semantic regularities,
and encodes them in the cosine distance between words.

For our experiments, we used GloVe~\citep{pennington2014glove} due to its
favorable performance characteristics.
GloVe works by leveraging the intuition that word-word co-occurrence
probabilities encode some form of meaning.
A classic example is the relationship between the word pair ``ice'' and ``steam''
and the word pair ``solid'' and ``gas.''
Gas and steam occur in the same sentence relatively frequently,
compared to the frequency with which the words gas and ice
occur in the same sentence.
Consequently, the following ratio is significantly less than $1$:
\[
\frac{\Pr(\text{gas}\mid\text{ice})}{\Pr(\text{gas}\mid\text{steam})}
\]
If, instead, we look at the frequency of sentences with both solid and ice
compared to the frequency of sentences with both solid and steam, we find the
opposite.
The ratio 
\[
\frac{\Pr(\text{solid}\mid\text{ice})}{\Pr(\text{solid}\mid\text{steam})}
\]
is much greater than $1$.
This signal is encoded into a large co-occurrence matrix.
GloVe then attempts to learn word vectors for which the dot-product of
two vectors is close to the logarithm of their probability of co-occurrence.



\begin{table*}[!tp]
\centering
\caption{Analogy Suite Details\label{Ta:ASUITE}}
\begin{tabular}{lllrrrr}
\toprule
Type    & Category         &  Representative Pair                                   & \# of Pairs  & Passing Tests   & Total Tests     & Accuracy         \\ \midrule
Calls   & 16 / 32          &  \cverb|store16/store32                  | & 18           & 246             & 306             & 80.39\%          \\
Calls   & Add / Remove     &  \cverb|ntb_list_add/ntb_list_rm         | & 9            & 72              & 72              & 100.0\%          \\
Calls   & Create / Destroy &  \cverb|device_create/device_destroy     | & 19           & 302             & 342             & 88.30\%          \\
Calls   & Enable / Disable &  \cverb|nv_enable_irq/nv_disable_irq     | & 62           & 3,577           & 3,782           & 94.58\%          \\
Calls   & Enter / Exit     &  \cverb|otp_enter/otp_exit               | & 12           & 122             & 132             & 92.42\%          \\
Calls   & In / Out         &  \cverb|add_in_dtd/add_out_dtd           | & 5            & 20              & 20              & 100.0\%          \\
Calls   & Inc / Dec        &  \cverb|cifs_in_send_inc/cifs_in_send_dec| & 10           & 88              & 90              & 97.78\%          \\
Calls   & Input / Output   &  \cverb|ivtv_get_input/ivtv_get_output   | & 5            & 20              & 20              & 100.0\%          \\
Calls   & Join / Leave     &  \cverb|handle_join_req/handle_leave_req | & 4            & 8               & 12              & 66.67\%          \\
Calls   & Lock / Unlock    &  \cverb|mutex_lock_nested/mutex_unlock   | & 53           & 2,504           & 2,756           & 90.86\%          \\
Calls   & On / Off         &  \cverb|b43_led_turn_on/b43_led_turn_off | & 19           & 303             & 342             & 88.60\%          \\
Calls   & Read / Write     &  \cverb|memory_read/memory_write         | & 64           & 3,950           & 4,032           & 97.97\%          \\
Calls   & Set / Get        &  \cverb|set_arg/get_arg                  | & 22           & 404             & 462             & 87.45\%          \\
Calls   & Start / Stop     &  \cverb|nv_start_tx/nv_stop_tx           | & 31           & 838             & 930             & 90.11\%          \\
Calls   & Up / Down        &  \cverb|ixgbevf_up/ixgbevf_down          | & 24           & 495             & 552             & 89.67\%          \\
Complex & Ret Check / Call &  \cverb|kzalloc_$NEQ_0/kzalloc           | & 21           & 252             & 420             & 60.00\%          \\
Complex & Ret Error / Prop &  \cverb|write_bbt_$LT_0/$RET_write_bbt   | & 25           & 600             & 600             & 100.0\%          \\ 
Fields  & Check / Check    &  \cverb|?->dmaops/?->dmaops->altera_dtype| & 50           & 2,424           & 2,450           & 98.94\%          \\ 
Fields  & Next / Prev      &  \cverb|!.task_list.next/!.task_list.prev| & 16           & 240             & 240             & 100.0\%          \\
Fields  & Test / Set       &  \cverb|?->at_current/!->at_current      | & 39           & 1,425           & 1,482           & 96.15\%          \\ \midrule
\textbf{Totals:} &         &                                            & \textbf{508} & \textbf{17,890} & \textbf{19,042} & \textbf{93.95\%} \\
\end{tabular}
\end{table*}



\rqsection{1}{useful}{Are Learned Vectors Useful?}

\Cref{rq:useful} asked whether vectors learned from abstracted
symbolic traces encode useful information.  We assess utility via
three experiments over word vectors.  Each of the
following subsections describes and interprets one experiment in
detail.

\newcommand{\Experiment}[1]{\subsection{Experiment \arabic{subsection}: #1}}

\Experiment{Code Analogies}%
\label{Se:CodeAnalogies}

An interesting aspect of word vectors is their ability to express
relationships between analogous words using simple math and cosine distance.
Encoding analogies is an intriguing byproduct of a ``good'' embedding and, as
such, analogies have become a common proxy for the
overall quality of learned word vectors.

\sloppy

No standard test suite for \emph{code} analogies exists, so we created
such a test suite using a combination of manual inspection and
automated search.
The test suite consists of twenty different categories,
each of which has some number of function pairs that have
been determined to be analogous.
For example, consider \cverb|mutex_lock_nested/mutex_unlock| and
\cverb|spin_lock/spin_unlock|; these are two pairs from the ``lock / unlock''
category given in \tableref{ASUITE}. We can construct an analogy by taking
these two pairs and concatenating them to form the analogy
``\cverb|mutex_lock_nested| is to \cverb|mutex_unlock| as \cverb|spin_lock| is to
\cverb|spin_unlock|.''
By identifying high-level patterns of behavior, and finding several
pairs of functions that express this behavior, we created a suite
that contains 19,042 code analogies.

\fussy

\tableref{ASUITE} lists our categories and the counts of available pairs,
along with a representative pair from each category.
\tableref{ASUITE} also provides accuracy metrics generated using
the vectors learned from what we will refer to as the ``baseline
configuration,''\footnote{The baseline configuration is described
  in more detail in \sectref{RQ2}, where it is also called \cref{config:baseline}.
}
which abstracts symbolic traces using all of the abstractions described in
in \sectref{ABS}.
We used a grid-search over hundreds of parameterizations to pick
hyper-parameters for our word-vector learner.
For the results described in this section, we used vectors of
dimension 300, a symmetric window size of 50, and a vocabulary-minimum
threshold of 1,000 to ensure that the word-vector learner only learns
embeddings for words that occur a reasonable number of times in the
 corpus of traces.
We trained for 2,000 iterations to give GloVe ample time to find good
vectors.

In each category, we assume that any two pairs of functions are
sufficiently similar to be made into an analogy.  More precisely, we
form a test by selecting two distinct pairs of functions %
$(\cA, \cB)$ and $(\cC, \cD)$ 
from the same category, and creating the triple
$(\cA, \cB, \cC)$ 
to give to an analogy solver that is equipped with
our learned vectors.  The analogy solver returns a vector $\cD'$, and
we consider the test passed if $\cD' = \cD$ and failed otherwise.
\Citet{levy-goldberg:2014:W14-16} present the following objective
to use when solving analogies with word-vectors:
%
\begin{align*}
\cD' = \argmax_{d \in \bV \setminus \{\cA,\cB,\cC\}} \;\cos(\,d,\cB\,) - \cos(\,d,\cA\,) + \cos(\,d,\cC\,)
\end{align*}

\subsubsubsection{Results}
The ``Accuracy'' column of \tableref{ASUITE} shows
that overall accuracy on the analogy suite is excellent.  Our
embeddings achieve greater than 90\% top-1 accuracy on thirteen out of
the twenty categories.  The learned vectors do the worst on the ``Ret
Check / Call'' category where the top-1 accuracy is only 60\%.
This category is meant to relate the checking of the return value
of a call with the call itself.
However, we often find that one function allocates memory, while a
different function checks for allocation success or failure.
For example, a wrapper function may allocate complex
objects, but leave callers to check that the allocation
succeeds.
Because our vectors are derived from intraprocedural traces,
it is sensible that accuracy suffers for interprocedural behaviors.

By contrast, our vectors perform extraordinarily well on the ``Ret
Error / Prop'' category (100\% top-1). This category represents cases where an
outer function (i) performs an inner call, (ii) detects that it has
received an error result, and (iii) returns (``propagates'') that error
result as the outer function's own return value.  Unlike for the ``Ret
Check / Call'' category, the nature of the ``Ret Error / Prop''
category ensures that both the check and the return propagation
can be observed in intraprocedural traces, without losing any information.

\Experiment{Simple Similarity}%
\label{Se:SimpleSimilarity}


\begin{figure}[tp]
  \centering
  \begin{tabular}{c}
\begin{lstlisting}
ret = new(/*...*/, &priv->bo);
if (!ret) {
  ret = pin(priv->bo, /*...*/);
  if (!ret) {
    ret = map(priv->bo);
    if (ret)
      unpin(priv->bo);
  }
  if (ret)
    ref(NULL, &priv->bo);
}
\end{lstlisting}
  \end{tabular}
  
  \setlength{\belowcaptionskip}{-18pt}
  \caption{Excerpt from \texttt{nv17\_fence.c}.  Names have been shortened to conserve space.\label{Fi:NVFENCE}}
\end{figure}



One of the most basic word-vector tasks is to ask for the $k$ nearest
vectors to some chosen vector (using cosine distance).  We expect the
results of such a query to return a list of relevant words from our
vocabulary.  Our similarity experiments were based on two types of queries:
(i) given a word, find the closest word, and
(ii) given a word, find the five closest words.

\subsubsubsection{Similar pairs}
We identified the single most similar word to
each word in our vocabulary $\bV$. This process produced thousands
of interesting pairs.  In the interest of space, we have selected
four samples which are representative of the variety of high-level relationships
encoded in our learned vectors\footnote{The artifact accompanying this paper includes
    a full listing of these pairs, ordered by cosine-similarity.}:

\begin{itemize}
\item \cverb|sin_mul| and \cverb|cos_mul|
\item \cverb|dec_stream_header| and \cverb|dec_stream_footer|
\item \cverb|rx_b_frame| and \cverb|tx_b_frame|
\item \cverb|nouveau_bo_new_$EQ_0| and \cverb|nouveau_bo_map|
  \footnote{In the following
text, and in \figref{NVFENCE}, we remove the
 \cverb|nouveau_bo_|
 prefix to conserve space.}
\end{itemize}

The last pair is of particular interest, because it expresses a complex pattern
of behavior that would be impossible to encode without some abstraction of the
path condition. The last pair suggests that there is a strong relationship
between the function \cverb|new| returning \cverb|0| (which signals a
successful call) and then the subsequent performance of some kind of map
operation with the \cverb|map| call.
To gain a deeper understanding of what the vectors are encoding, we searched
for instances of this behavior in the original source code.
We found several instances of the pattern shown in \figref{NVFENCE}.

The code in \figref{NVFENCE} raise a new question: why isn't
\cverb|pin| more closely related to
\cverb|new_$EQ_0|?  We performed additional similarity queries to gain
a deeper understanding of how the learned vectors have encoded the
relationship between \cverb|new|, \cverb|pin|, and \cverb|map|.

First, we checked to see how similar \cverb|pin| is to
\cverb|new_$EQ_0|. We found that \cverb|pin| is the fourth-most related word to
\cverb|new_$EQ_0|, which suggests that a relationship does exist, but that the
relationship between \cverb|new_$EQ_0| and \cverb|pin| is not as strong as the
one between \cverb|new_$EQ_0| and \cverb|map|. Looking back at the code snippet
(and remembering that several more instances of the same pattern can be found in
separate files), we are left with the fact that \cverb|pin| directly follows from
the successful \cverb|new|.  Therefore, intuition dictates that \cverb|pin|
should be more strongly related to \cverb|new| than \cverb|map|. The
disagreement between our intuition and the results of our word-vector queries
motivated us to investigate further.

By turning to the traces for an answer, we uncovered a more complete picture.
In 3,194 traces, \cverb|new| co-occurs with \cverb|pin|.
In 3,145 traces, \cverb|new| co-occurs with \cverb|map|.
If we look at traces that do \emph{not} contain a call to \cverb|new|,
there are $11,354$ traces that have no call to \cverb|new|, but still
have a call to \cverb|pin|.
In contrast, only $352$ traces have no call to \cverb|new|, but still have a call
to \cverb|map|.
Finally, we have a definitive answer to the encoding learned by the vectors:
it is indeed the case that \cverb|new| and \cverb|map| are more related in our corpus of traces,
because almost every time a call to \cverb|map| is made,
a corresponding call to \cverb|new| precedes it.
Our intuition fooled us, because the snippets of source code only revealed a
partial picture.

\subsubsubsection{Top-5 similar words and the challenge of prefix dominance}
Another similarity-based test is to take a word
and find the top-$k$ closest words in the learned embedding space. Ideally, we'd
see words that make intuitive sense. For the purpose of evaluation, we picked
two words: \cverb|affs_bread|, a function in the AFS file system that
reads a block, and \cverb|kzalloc|, a memory allocator.
For each target word, we evaluated the top-5 most similar words for relevance.
In the process, we also uncovered an interesting challenge when
learning over path-sensitive artifacts, which we call \textit{prefix
dominance}.

Our corpus of symbolic traces can be thought of as a corpus of execution trees.
In fact, in the implementation of our trace generator, the traces only exist at
the very last moment.
Instead of storing traces, we store a tree that encodes, without unnecessary
duplication, the information gained from symbolically executing a procedure.
If we think about the dataset of traces as a dataset of trees (each of which holds many
traces that share common prefixes), we begin to see that learning word vectors
from traces is an approximation of learning directly from the execution
trees.

The approximation of trees by traces works, in the sense that we
can use the traces to learn meaningful vectors, but the approximation
is vulnerable to learning rare behaviors that exist at the beginning
of a procedure whose trace-tree has many nested branches.
These rare behaviors occur only once in the original procedure text
and corresponding execution tree, but are replicated many times in the traces.
In a procedure with significant branching complexity, a single
occurrence of rare behavior can easily overwhelm any arbitrary number
of occurrences of expected behavior.

\sloppy

In \tableref{TOP5}, we see two words, \cverb|affs_bread| and
\cverb|kzalloc|, and the five most similar words to each of them.
Word similarity has captured many expected relationships.  For
example, the fact that \cverb|kzalloc| is most commonly checked to be
non-null
(\cverb|kzalloc_$NEQ_0|) and then also \cverb|kfree|d is what we
would expect, given the definition of an allocator.  Similarly, we can
see that \cverb|affs_bread| is also checked to
be non-null, checksummed, freed, released, etc.
However, in addition to these expected relationships,
the last three entries for \cverb|kzalloc| seem out of place.
These unexpected entries are present in the top-5 answer
because of prefix dominance.

\fussy

\begin{table}[t]
\centering
\caption{Top-5 closest words to \texttt{affs\_bread} and \texttt{kzalloc} \label{Ta:TOP5}}
\begin{tabular}{ll}
\toprule
\textbf{\texttt{\;affs\_bread}} & \textbf{\texttt{\;kzalloc}}       \\
\midrule
\cverb|affs_bread_$NEQ_0  |   & \cverb|kzalloc_$NEQ_0             | \\
\cverb|affs_checksum_block|   & \cverb|kfree                      | \\
\cverb|AFFS_SB            |   & \cverb|_volume                    | \\
\cverb|affs_free_block    |   & \cverb|snd_emu10k1_audigy_write_op| \\
\cverb|affs_brelse        |   & \cverb|?->output_amp              | \\
\bottomrule
\end{tabular}
\end{table}



We searched our traces for places where \cverb|kzalloc| and the three
unexpected entries in the table co-occur. We found one function with
5,000 paths (5,000 being our ``budget'' for the number of traces we
are willing to generate via symbolic execution for a single procedure), of which 4,999 have
several instances of the pattern \cverb|kzalloc| followed by
\cverb|snd_emu10k1_audigy_write_op|. This one function, with its
multitude of paths, overwhelms our dataset, and causes the word vectors
to learn a spurious relationship.  Prefix dominance also explains the
strong associations between \cverb|kzalloc| and \cverb|_volume| and
\cverb|?->output_amp|.

On the other hand, \cverb|affs_bread| is relatively unaffected by
prefix dominance.  Examining our traces for the \cverb|affs| file
system that contains this function, we found that no procedures had an
overwhelming number of paths.  Therefore, we never see an overwhelming
number of \cverb|affs_bread| usage patterns that are rare at the
source level but common in our set of traces.

\Experiment{Queries Via Word-Vector Averaging}

Word vectors have the surprising and useful ability to encode meaning
when averaged~\citep{Le:2014:DRS:3044805.3045025,kenter2016siamesecbow}. 
We devised a test to see if our learned vectors are
able to leverage this ability to capture a relationship between
allocation failure and returning \cverb|-ENOMEM|.

To understand whether our word vectors are capable of answering such a
high-level question, we evaluated their performance on increasingly
targeted queries (represented by averaged vectors).  Each query was
restricted to search only for words in the subspace of the embedding
space that contains kernel error-codes.
(Narrowing to the subspace of error codes ensures that we are only
looking at relevant words, and not at the whole vocabulary.)

\subsubsubsection{Results}
We identified twenty different functions that act as allocators in
the Linux kernel.

First, for each such allocator, we took its word vector $\cA$, and
queried for the closest vector to $\cA$ (in the subspace of error codes).
This method found the correct error code only twice out of twenty
tests (i.e., 10\% accuracy).

Second, we asked for the vector closest to an average vector that
combined the vector for the allocator $\cA$ of interest and the vector
$\vERR$ for a generic error:\footnote{The
  \cverb|$ERR| word is added to any trace that returns either (i) the
  result of an \cverb|ERR_PTR| call, or (ii) a constant less than zero that
  is also a known error code.
  Consequently, a vector $\vERR$ is learned for the word \cverb|$ERR|.
}
$(\cA + \vERR) / 2$. 
This query found the correct \cverb|ENOMEM| code fourteen times out of
twenty (i.e., 70\% accuracy).

Third, instead of averaging the allocator's $\cA$ vector with $\vERR$,
we tried averaging $\cA$ with the vector for the special \cverb|$END|
token that signals the end of a trace.
Seeking the error code closest to $(\cA + \vEND) / 2$ found the
correct result for sixteen of twenty test cases (i.e., 80\% accuracy).
The fact that this method outperforms our previous query reveals that
the call to an allocator being near the end of a trace is an even
stronger signal than the \cverb|$ERR| token.

\sloppy

Finally, we mixed the meaning of the allocator, the error token, and
the end-of-trace token by averaging all three:
$(\cA + \vERR + \vEND) / 3$. 
The error code whose vector is closest to this query is the correct
\cverb|ENOMEM| code for eighteen of the twenty tests (i.e., 90\%
accuracy).
The steadily increasing performance indicates that targeted queries
encoded as average word vectors can indeed be semantically meaningful.

\fussy

The effectiveness of these queries, and the results from
\sectrefs{CodeAnalogies}{SimpleSimilarity}, support a positive answer
to \cref{rq:useful}: learned vectors do encode useful information
about program behaviors.



\rqsection{2}{best}{Ablation Study}

In this section, we present the results of an ablation study
to isolate the effects that different sets of abstractions
have on the utility of the learned vectors.
We used the benchmark suite of 19,042 code-analogies from \sectref{RQ1}
to evaluate eight different configurations.
We scored each configuration according to the number of analogies correctly
encoded by the word vectors learned for that configuration (i.e.,
we report top-1 results).

In addition to the baseline configuration from
\sectref{CodeAnalogies}, we partitioned the abstractions into six
classes\footnote{ Except for
  \lstinline{Called}, which was used in all configurations.  } and
generated six new embeddings, each with one of the six abstraction
classes excluded.  We also used one more configuration in which
\emph{stop words} were included.  In natural language processing, stop
words are words that are filtered out of a processing toolchain.
Sometimes these are the most common words in a language, but any group
of words can be designated as stop words for a given application.  In
our context, stop words are function names that occur often, but add
little value to the trace.  Examples are \cverb|__builtin_expect| and
automatically generated \cverb|__compiletime_assert|s.

We evaluated the following eight configurations:
\begin{enumerate}
\item \textbf{baseline}: all abstractions from
  \sectref{ABS};\label{config:baseline}

\item baseline without \lstinline{ParamTo} and
  \lstinline{ParamShare};\label{config:no-param}

\item baseline without \lstinline{RetEq}, \lstinline{RetNeq}, etc.;

\item baseline without \lstinline{AccessPathStore} and
  \lstinline{AccessPathSensitive};\label{config:no-access}

\item baseline without \lstinline{PropRet}, \lstinline{RetError},
  and \lstinline{RetConst};

\item baseline without \lstinline{Error};

\item baseline without \lstinline{FunctionStart} and
  \lstinline{FunctionEnd}; and

\item baseline with stop words included.
\end{enumerate}


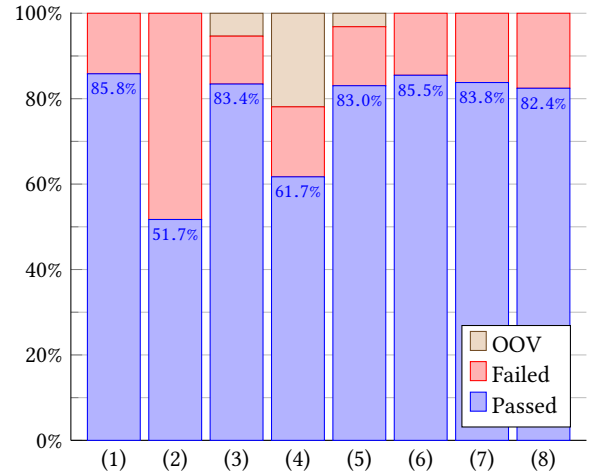
\begin{figure}
\begin{tikzpicture}
  \begin{axis}[
    trinary plot={x}{y},
    bar width=20pt,
    legend cell align=left,
    legend pos=south east,
    reverse legend,
    xticklabel={(\pgfmathprintnumber{\tick})},
  ]

  \pgfplotstableread{fig/ablation.csv}\data
  \addsumcolumn{\data}

  \addplot+[percentage nodes near coords={below, font=\footnotesize}] table[percentage series={y}{Passed}] \data;
  \addplot+ table[percentage series={y}{Failed}] \data;
  \addplot+ table[percentage series={y}{OOV}] \data;

  \legend{Passed, Failed, OOV}

  \end{axis}
\end{tikzpicture}
\setlength{\belowcaptionskip}{-10pt}
\vspace{-1.5ex}
\caption{Ablation study: top-1 analogy results for eight
  configurations (baseline \labelcref{config:baseline} with up to one
  individual abstraction class removed).  The vocabulary minimum was
  0, and the number of training iterations was 1,000.\label{Fi:ABL}}
\end{figure}


\newcommand{\bartext}[2]{{%
  \setlength{\fboxsep}{0pt}%
  \colorbox{#1!30!white}{\vphantom{Xy}#2}}}

\figref{ABL} compares the accuracy of for these eight configurations.
\bartext{blue}{Blue} bars indicate the number of tests in the analogy
suite that passed; \bartext{red}{red} indicates tests that failed; and
\bartext{brown}{brown} indicates out-of-vocabulary (OOV) tests.
\Cref{config:no-access} had the most out-of-vocabulary tests; in this
configuration, we do not have words like \cverb|!->next| and
\cverb|!->prev|, which leaves several portions of the analogy suite
essentially unanswerable.  Thus, we count out-of-vocabulary tests as
failed tests.

To create a fair playing field for evaluating all eight
configurations, we chose a single setting for the hyper-parameters
that were used when learning word vectors.  We reduced the threshold
for how often a word must occur before it is added to the vocabulary
from 1,000 to 0.  The latter parameter, which we refer to as the
\textit{vocabulary minimum}, significantly impacts performance by
forcing the word-vector learner to deal with thousands of rarely-seen
words.  To understand why we must set the vocabulary minimum to zero,
effectively disabling it, consider the following example trace:
\lstinline{Called(foo), ParamShare(foo, bar), Called(bar)}.  In
\cref{config:no-param}, where we ignore \lstinline{ParamShare}, we
would encode this trace as the sentence \cverb|foo bar|. In
\cref{config:baseline}, this same trace is encoded as %
\cverb|foo foo bar|.  The fact that some abstractions can influence
the frequency with which a word occurs in a trace corpus makes any
word-frequency-based filtering counterproductive to our goal of
performing a fair comparison.

We also lowered the number of training iterations from 2,000 to 1,000
to reduce the resources required to run eight separate configurations
of our toolchain.  (These changes are responsible for the
change in the top-1 accuracy of the baseline configuration from 93.9\%
in \tableref{ASUITE} to 85.8\% in \figref{ABL}.)

In \figref{ABL}, one clearly sees that \cref{config:no-param} (the one
without any dataflow-based abstractions) suffers the worst performance
degradation.  \Cref{config:no-access}, which omits access-path-based
abstractions, has the second-worst performance hit.  These results
indicate that dataflow information is critical to the
quality of learned vectors.  This conclusion further confirms findings
by \citet{DBLP:journals/corr/abs-1711-00740} regarding the importance
of dataflow information when learning from programs.

\figref{ABL} also reveals that removing ``state'' abstractions
(\lstinline{RetEq}, \lstinline{RetNeq}, etc. and \lstinline{Error})
has little effect on quality.  However, these abstractions still add
useful terms to our vocabulary, and thereby enlarge the set of
potentially answerable questions.  Without these abstractions, some of
the questions in \sectref{RQ1} would be unanswerable.

These results support the following answer to \cref{rq:best}:
dataflow-based abstractions provide the greatest benefit to
word-vector learning.  These abstractions, coupled with
access-path-based abstractions, provide sufficient context to let a
word-vector learner create useful embeddings.  Adding abstractions
based on path conditions (or other higher-level concepts like
\lstinline{Error}) adds flexibility without worsening the quality of
the learned vectors.  Therefore, we recommend including these
abstractions, as well.



\rqsection{3}{svs}{Syntactic Versus Semantic}

Now that we have seen
the utility of the generated corpus for word-vector learning (\sectref{RQ1})
and the interplay between the abstractions we use (\sectref{RQ2}),
we compare our recommended \cref{config:baseline} from \sectref{RQ2} with a simpler
syntactic-based approach.

We explored several options for a syntactic-based approach against
which to compare.
In trying to make a fair comparison, one difficulty that arises
is the amount of data our toolchain produces to use for the
semantics-based approach.
If we were to compare \cref{config:baseline} against an approach based
on ASTs or tokens, there would be a large disparity between the
paucity of data available to the AST/token-based approach compared to
the abundance of data available to the word-vector learner:
an AST- or token-based approach would only have one data point
per procedure, whereas the path-sensitive artifacts gathered
using \cref{config:baseline} provide the word-vector learner with
hundreds, if not thousands, of data points per procedure.

To control for this effect and avoid such a disparity, we instead compared
\cref{config:baseline} against a configuration of our toolchain
that uses only ``syntactic'' abstractions---i.e., abstractions that
can be applied without any information obtained from symbolic execution.
Thus, the syntactic abstractions are:
\begin{itemize}
  \item \lstinline{FunctionStart} and \lstinline{FunctionEnd},
  \item \lstinline[mathescape]{AccessPathStore($\mathit{path}$)}, and
  \item \lstinline[mathescape]{Called($\mathit{callee}$)}.
\end{itemize}
The rest of our abstractions use deeper semantic information, such as
constant propagation, dataflow information, or the path condition
for a given trace.


\begin{figure}
  \begin{tikzpicture}
    \begin{axis}[
      height=1.3in,
      width=\axisdefaultwidth,
      trinary plot={y}{x},
      bar width=15pt,
      legend pos=north west,
      legend style={
        anchor=south west,
        legend columns=-1,
      },
      table/y index=0,
      symbolic y coords={
        Syntactic,
        Semantic,
      },
      enlarge y limits={
        true,
        value=.7,
      },
      tick label style={font=\small},
      xticklabel={\pgfmathprintnumber{\tick}\%},
    ]

    \pgfplotstableread{fig/rq3.csv}\data
    \addsumcolumn{\data}

    \addplot+[percentage nodes near coords=left] table[percentage series={x}{Passed}] \data;
    \addplot+ table[percentage series={x}{Failed}] \data;
    \addplot+ table[percentage series={x}{OOV}] \data;

    \legend{Passed, Failed, OOV}

    \end{axis}
  \end{tikzpicture}
  \setlength{\belowcaptionskip}{-15pt}
  \caption{Top-1 analogy results for syntactic versus semantic abstractions.
    (The vocabulary minimum was 0, and the number of training iterations was 1,000.)
    \label{Fi:RQ3}
  }
\end{figure}
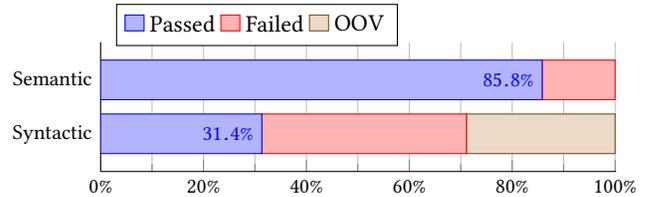


Using only the syntactic abstractions, we generated a
corpus of traces, and then learned word vectors from the corpus.
We compared the newly learned word vectors to the ones obtained
with \cref{config:baseline}.
\figref{RQ3} clearly shows that semantic abstractions are crucial to
giving the context necessary for successful learning.
Even if we assess performance using only the analogies that are
in-vocabulary for the syntactic-based approach, we find that
the syntactic-based approach achieves only about 44\% accuracy,
which is \textit{about half} the accuracy of vectors learned
from (mainly) semantic abstractions.

These results support an affirmative answer to \cref{rq:svs}:
abstracted traces that make use of semantic information obtained via
symbolic execution provide more utility as the input to a word-vector
learner than abstracted traces that use only syntactic information.



\rqsection{4}{downstream}{Use in Downstream Tasks}

\Cref{rq:downstream} asks if we can utilize our pre-trained word-vector
embeddings on some downstream task. 

To address this question, we selected a downstream task that models 
bug finding, repair, and code completion in a restricted domain: error-code misuse. 
We chose error-code misuse because it allows us to apply supervised learning. 
Because there are only a finite number of common error codes in the Linux kernel, we can 
formulate a multi-class labeling problem using traces generated via 
our toolchain and our pre-trained word-vector embeddings.

To build an effective error-code-misuse model, 
we gathered a collection of failing traces (traces in which the \cverb|$ERR| token occurs).
We then constructed a dataset suitable for supervised learning as follows:
we took each trace from configuration (2)\footnote{The dataflow abstractions 
present in (1) were created to aid word-vector learners; for this experiment, we use configuration (2) to 
exclude those abstractions.} and removed the last three abstract tokens, namely,
\cverb|$ERR|, \cverb|$RET_E*|, and \cverb|$END|;\footnote{
  We exclude traces that included the \cverb|$RET_PTR_ERR| token because these traces
  do not have an associated error code.
}
we used the \cverb|$RET_E*| token as the label for the trimmed trace. We sampled 
a subset of 20,000 traces from this large trace collection to use for training our model. 

This dataset is a good starting point, but feeding it to a machine-learning 
technique that accepts fixed-length inputs requires further processing.
To preprocess the data, we kept only the last 100 tokens in each trace.
We then took the trimmed traces, and used our learned
word-vector embedding to transform each sequence of words into a
sequence of vectors (of dimension 300).
If, originally, a trace had fewer than 100 tokens, we padded the beginning of the trace
with the zero vector.
We paired each of the trimmed and encoded traces with its
label (which we derived earlier).
Finally, to complete the preprocessing of the dataset we attached a one-hot encoding of the label. 

To collect a challenging test set to evaluate our learned model, we turned to 
real bug-fixing commits applied to the Linux kernel. We searched for commits that
referenced an ``incorrect return'' in their description.
In addition, we leveraged \citeauthor{Min:2015:CSC:2815400.2815422}'s \citep{Min:2015:CSC:2815400.2815422} list of incorrect return
codes fixed by their JUXTA tool.  
Next, we generated abstracted symbolic traces both before applying the fixing commit and after. 
Finally, we kept the traces generated before applying the fix that, after the fix, 
had changed only in the error code returned. Using this process,
we collected 68 traces---from 15 unique functions---that had been patched
to fix an incorrect return code.


Using the preprocessed dataset, we trained a model to predict the error code
that each trace should return. We used a recurrent neural network with 
long short-term memory (LSTM) \citep{Hochreiter:1997:LSM:1246443.1246450}. 
We evaluated the trained model, using our test set, 
in two different ways:

\begin{enumerate}
\item \textit{Bug Finding:} we use our learned model to predict the three 
most likely error codes for each trace in our test set. If a given trace 
initially ended in the error code \cverb|A|, but was patched to return the error
code \cverb|B|, we check to see if the incorrect \cverb|A| error code is
absent from our model's top-3 predictions.

\item \textit{Repair / Suggestion:} we again use the learned model to 
predict the three most likely error codes for each trace in the test set. 
This time, we determine the fraction of traces for which the correct
error code (i.e., \cverb|B|) is present in the top-3 prediction made by the model.
\end{enumerate}

In evaluation (1), we found that the learned model identified an 
incorrect error code in 57 of our 68 tests. This result is promising, because it suggests
that there is enough signal in the traces of encoded vectors to make good 
predictions that could be used to detect bugs early.

In evaluation (2), we observed that the learned model had a top-3 accuracy 
of 76.5\%, meaning that the correct error code is among our top
suggested fixes for more than three fourths of the buggy traces.
This result is a strong indicator that the learned vectors and
abstracted symbolic traces are rich enough to make high-level predictions 
that could be used to augment traditional IDEs with predictive capabilities. 
Such a feature could operate like autocomplete, but with an awareness
of what other contributors
have done and how their (presumably correct) code should influence new 
contributions. This feature would be similar to the existing applications of statistical 
modeling to programming tasks such as autocompletion 
\citep{Allamanis:2015:SAM:2786805.2786849, Raychev2015, Bielik:2016:PPM:3045390.3045699, Nguyen:2013:SSL:2491411.2491458, Nguyen2012}.

These results support an affirmative answer to \cref{rq:downstream}: our 
pre-trained word-vector embeddings can be used successfully on downstream 
tasks. These results also suggest that there are many interesting applications
for our corpus of abstracted symbolic traces. Learning from these traces to 
find bugs, detect clones, or even suggest repairs, are all within the realm
of possibility.



\section{Threats to Validity\label{Se:THREATS}}

There are several threats to the validity of our work. 

We leverage a fast, but imprecise, symbolic-execution engine. It is possible
that information gained from the detection of infeasible paths and the
use of a memory model would improve the quality of our learned vectors.
In addition, it is likely that a corpus of interprocedural traces would
impact our learned vectors.

We chose to focus our attention on the Linux kernel. It is possible
that learning good word-embeddings using artifacts derived from the
Linux kernel does not translate to learning good word-embeddings for
programs in general.
To mitigate this risk, we maximized the amount of diversity in the
ingested procedures by ingesting the Linux kernel with all modular and
optional code included.

Our analogies benchmark and the tests based on word-vector averaging
are only proxies for meaning, and, as such, only serve as an indirect
indicator of the quality of the learned word vectors.
In addition, we created these benchmarks ourselves, and thus there is
a risk that we introduced bias into our experiments.
Unfortunately, we do not have benchmarks as extensive as those created throughout
the years  in the NLP community.
Similar to  \citet{DBLP:journals/corr/abs-1301-3781}, we hope that our introduction 
of a suitable benchmark will facilitate comparisons between different learned embeddings 
in the future.


\section{Related Work\label{Se:RW}}

Recently, several techniques have leveraged learned embeddings for artifacts
generated from programs.
\Citet{Nguyen2017,Nguyen:2016:MAE:2889160.2892661} leverage word embeddings
(learned from ASTs) in two domains to facilitate translation from Java to C\#.
\Citet{Pradel2017} use embeddings (learned from custom
tree-based contexts built from ASTs) to bootstrap anomaly detection against a
corpus of JavaScript programs.  \Citet{Gu:2016:DAL:2950290.2950334} leverage an
encoder/decoder architecture to embed whole sequences in their \textsc{DeepAPI}
tool for API recommendation. \textsc{API2API}
by \citet{7886921} also leverages word embeddings, but it learns the
embeddings from API-related natural-language documents instead of an artifact
derived directly from source code.

Moving toward more semantically rich embeddings, \citet{DBLP:journals/corr/abs-1802-07779}
leverage labeled pushdown systems to generate rich traces which they use to learn 
function embeddings.  They apply these embeddings to find function synonyms, which can be used to improve
traditional specification mining techniques. \Citet{Alon:2018:GPR:3192366.3192412} learn
from paths through ASTs to produce general representations of programs; in \citep{DBLP:journals/corr/abs-1803-09473} 
they expand upon this general representation by leveraging attention mechanisms. \Citet{ncc} utilize 
an intermediate representation (IR) to produce embeddings of programs that are learned from both control flow and 
data flow information.

Venturing into general program embeddings, there are several recent
techniques that approach the problem of embedding programs (or, more
generally, symbolic-expressions/trees) in unique ways. Using
input/output pairs as the input data for learning,
\citet{Piech:2015:LPE:3045118.3045235} and
\citet{DBLP:journals/corr/ParisottoMSLZK16} learn to embed whole
programs. Using sequences of live variable values, \citet{DBLP:journals/corr/abs-1711-07163}
produce embeddings to aid in program repair tasks. \Citet{DBLP:journals/corr/abs-1711-00740} learn to embed
whole programs via Gated Graph Recurrent Neural Networks (GG-RNNs)
\citep{DBLP:journals/corr/LiTBZ15}.  \Citet{Allamanis2016a} approach
the more foundational problem of finding continuous representations of
symbolic expressions. \Citet{AAAI1611775} introduce tree-based
convolutional neural networks (TBCNNs), another model for embedding
programs.  \Citet{Peng:2015:BPV:2978872.2978936} provide 
an AST-based encoding of programs with the goal of facilitating
deep-learning methods. \Citet{DBLP:journals/corr/abs-1709-06182} give a
comprehensive survey of these techniques, and many other applications
of machine learning to programs.

We are not aware of any work that attempts to embed traces generated
from symbolic execution. On the contrary, \citet{Fowkes:2016:PPA:2950290.2950319} warn of
possible difficulties learning from path-sensitive artifacts.
We believe that our success in using symbolic traces as the input
to a learner is due to the addition of path-condition and dataflow
abstractions---the extra information helps to ensure that a complete
picture is seen, even in a path-sensitive setting.

In the broader context of applying statistical NLP techniques to programs, there 
has been a large body of work using language models to understand programs  
\citep{Hindle2012a,Raychev:2014:CCS:2594291.2594321,Nguyen:2015:GSL:2818754.2818858,
Allamanis:2015:BMS:3045118.3045344,Allamanis:2014:LNC:2635868.2635883};
to find misuses~\citep{Murali:2017:BSL:3106237.3106284,Wang:2016:BBD:2970276.2970341};
and to synthesize expressions and code snippets
\citep{Gvero:2015:SJE:2814270.2814295,Raghothaman:2016:SSI:2884781.2884808}.




\section{Experimental Artifacts}%
\label{Se:Artifact}

\sloppy

Our \texttt{c2ocaml} tool, which performs a source-to-source 
transformation during the compilation of C projects (to generate
inputs to our lightweight symbolic execution engine) is available 
at \href{https://github.com/jjhenkel/c2ocaml}{\texttt{https://github.com/jjhenkel/c2ocaml}}.

\fussy

Our lightweight symbolic execution engine, \texttt{lsee}, is also 
available at \href{https://github.com/jjhenkel/lsee}{\texttt{https://github.com/jjhenkel/lsee}}.

\sloppy

Additionally, we provide tools to demonstrate our experiments at 
\href{https://github.com/jjhenkel/code-vectors-artifact}{\texttt{https://github.com/jjhenkel/code-vectors-artifact}} \citep{jordan_henkel_2018_1297689}.
This artifact allows the user to run our toolchain end-to-end on a smaller 
open-source repository (Redis). The artifact uses pre-built Docker~\citep{Merkel:2014:DLL:2600239.2600241} images to avoid 
complex installation requirements. Our raw data (two sets of learned vectors and a full collection 
of abstracted symbolic traces) are also included in this artifact.

\fussy


\vspace{-5pt}
\section{Conclusion\label{Se:CONC}}

The expanding interest in treating programs as data to be fed to 
general-purpose learning algorithms has created a need for methods 
to efficiently extract, canonicalize, and embed artifacts
derived from programs.
In this paper, we described a toolchain for efficiently extracting
program artifacts;
a parameterized framework of abstractions for canonicalizing
these artifacts;
and an encoding of these parameterized embeddings in a format that can
be used by off-the-shelf word-vector learners. 

Our work also provides a new benchmark to probe the quality of word-vectors
learned from programs.
Our ablation study used the benchmark to provide insight about which
abstractions contributed the most to our learned word vectors.
We also provided evidence that (mostly) syntactic abstractions are
ill-suited as the input to learning techniques.
Lastly, we used these tools and datasets to learn a model of a
specific program behavior (answering the question, ``Which error is a trace likely to return?''),
and applied the model in a case study to confirm actual bugs found
via traditional static analysis.


\begin{acks}
  This research was supported, in part, by a gift from
  \grantsponsor{Batra}{Rajiv and Ritu
    Batra}{http://www.cs.wisc.edu/awards/scholarships.lm.html}; by
  \grantsponsor{AFRL}{AFRL}{http://www.wpafb.af.mil/afrl/} under
  \grantsponsor{DARPA}{DARPA}{https://www.darpa.mil/} MUSE award
  \grantnum{DARPA}{FA8750-14-2-0270} 
  and DARPA STAC award \grantnum{DARPA}{FA8750-15-C-0082}; 
  by \grantsponsor{ONR}{ONR}{https://www.onr.navy.mil/} under grant
  \grantnum{ONR}{N00014-17-1-2889}; 
  by \grantsponsor{NSF}{NSF}{https://www.nsf.gov/} under grants
  \grantnum{NSF}{CCF-1318489}, \grantnum{NSF}{CCF-1420866}, and
  \grantnum{NSF}{CCF-1423237}; and by the
  \grantsponsor{UWVC}{UW--Madison 
    Office of the Vice Chancellor for Research and Graduate
    Education}{https://research.wisc.edu/} with funding from the
  \grantsponsor{WARF}{Wisconsin Alumni Research
    Foundation}{https://www.warf.org/}.  Any opinions, findings, and
  conclusions or recommendations expressed in this publication are
  those of the authors, and do not necessarily reflect the views of
  the sponsoring agencies.
\end{acks}



\bibliographystyle{acm/ACM-Reference-Format}
\bibliography{bib/henkel,bib/programs-as-data}

\end{document}